\PassOptionsToPackage{numbers,sort&compress}{natbib}
\documentclass[preprint,12pt]{elsarticle}
\usepackage[total={6.25in, 8.2in}]{geometry}
\usepackage{amssymb}
\usepackage{amsmath}
\usepackage{caption} 
\usepackage{booktabs}

\usepackage{setspace}
\makeatletter
\def\ps@pprintTitle{%
   \let\@oddhead\@empty
   \let\@evenhead\@empty
   \def\@oddfoot{\hfill\thepage\hfill}
   \let\@evenfoot\@oddfoot}
\makeatother

\begin{document}

\begin{frontmatter}

\title{Atomistic mechanism of corrosion-induced grain boundary migration in NiCr alloys in molten FLiNaK}

\author[1]{Sadia Khan}
\author[1]{Hamdy Arkoub}
\author[1]{Miaomiao Jin\corref{cor}}
\affiliation[1]{organization={Department of Nuclear Engineering, The Pennsylvania State University},
            city={University Park},
            state={PA},
            postcode={16802},            
            country={USA}}   
\cortext[cor]{Corresponding author: mmjin@psu.edu}

\begin{abstract}
  
Corrosion of Ni--Cr structural alloys in molten fluoride salts is a persistent material degradation problem, yet the atomistic role of grain boundaries in this process remains poorly understood. Here we use reactive molecular dynamics to investigate corrosion of NiCr alloys in molten FLiNaK across four representative grain boundaries ($\Sigma3(111)$, $\Sigma11(113)$, $\Sigma5(012)$, and $\Sigma5(013)$) and corresponding bulk surfaces. Surface crystallography controls the initial dissolution stage, while grain boundary character governs the spatial localization and longer-time evolution of corrosion. We further identify a corrosion--driven grain boundary migration mechanism in which fluorine localization, preferential chromium dissolution, and vacancy--mediated mobility together drive interfacial motion away from the dealloyed region. The coherent $\Sigma3(111)$ boundary suppresses these processes, indicating low--energy special boundaries as targets for grain boundary engineering of corrosion--resistant Ni--Cr alloys.

\end{abstract}


\end{frontmatter}

\section{Introduction}

The corrosion of Ni--Cr structural alloys in molten fluoride salts remains a long--standing and technologically significant materials degradation problem, relevant to a range of high--temperature energy and processing applications \cite{Robin2022, flinak, flinak2}. Among molten fluorides, the eutectic LiF--NaF--KF (FLiNaK) combines favorable thermophysical properties with chemical stability at elevated temperatures \cite{Guo2018}, yet remains aggressive toward most structural alloys \cite{dealloy}. Ni--Cr alloys are widely regarded as leading structural candidates for service in molten fluoride environments because they combine high-temperature strength with comparatively favorable thermodynamic stability in fluoride media \cite{Zheng2018}. Even these alloys, however, degrade through preferential Cr dissolution, surface dealloying, and intergranular attack, producing Cr--depleted and Ni--enriched surfaces that compromise long--term integrity \cite{chan2023, chan2024, arkoub2025, dealloy}. Understanding the atomistic basis of this degradation, particularly the role of microstructural features such as grain boundaries (GBs), is essential for the rational design of corrosion--resistant alloys for molten salt environments.

The corrosion of Ni--Cr alloys in molten fluoride salts is governed by the coupled effects of surface chemistry and microstructural pathways. The underlying mechanism in NiCr/fluoride systems involves fluorine adsorption at the alloy surface, which weakens Cr--Ni metallic coordination and drives selective Cr removal into the salt \cite{arkoub2024, Yin, Cuilan}. GBs are critical corrosion pathways. GB character has long been recognized as a key control on intergranular corrosion susceptibility across a range of alloy systems and environments, with low--energy coherent boundaries more resistant than high--angle random boundaries, and with GB-network connectivity controlling intergranular propagation \cite{Gertsman2001, Bettayeb2018, Ebrahimi, Hu2011, Barr2018, twingb}. Similar trends have been documented in Ni--based alloys exposed to molten fluorides, where pronounced intergranular attack and preferential GB dealloying into the alloy have been characterized experimentally \cite{igc, Gu2022, Yang2020, Zhou2020, chan2024}, and first--principles calculations on a $\Sigma5$ GB in NiCr/FLiNaK have further shown that GB sites enhance fluorine binding and lower the energy barrier for Cr dissolution relative to bulk \cite{ARKOUB2025113903}. 

Beyond serving as fast-diffusion and salt infiltration pathways, GBs are themselves dynamically restructured during molten salt corrosion, and this dynamic behavior has been explicitly interpreted through the framework of diffusion-induced grain boundary migration (DIGM). Yang et al.~\cite{yang2023} documented a one-dimensional "wormhole" corrosion mode in Ni--20Cr exposed to molten salt and identified a vacancy-supersaturated DIGM zone that serves as the precursor to wormhole formation. Teng et al.~\cite{teng2025} likewise attributed corrosion--induced voids and GB waviness in molten--salt--corroded Ni--20Cr directly to DIGM. Recent experimental work by Walter et al. \cite{walter2026} has further demonstrated that DIGM plays a central role in molten chloride salt corrosion of Ni–-30Cr alloys by coupling GB migration with preferential Cr transport and depletion. The classical DIGM framework originates from vapor--solid and solid-state systems, where solute uptake along GBs establishes compositional gradients that drive boundary migration \cite{gbshift3, gbshift2, gbshift4}. This concept has subsequently been extended to corrosion phenomena such as intergranular oxidation of Ni-based alloys in pressurized water, where inward oxygen diffusion drives GB migration and produces GB--aligned oxide products \cite{gbshift}. Recent atomistic work on Alloy 690 has further shown that DIGM in Ni-Cr alloys depends on GB character, with the driving force set by coupled solute diffusivity and segregation effects \cite{jiang2026}. The molten-salt situation, however, presents a qualitatively distinct case: the GB is exposed to continuous solute removal through dealloying while remaining in contact with a redox-active molten salt.

Despite this phenomenological picture, the atomistic-scale understanding of how GBs govern corrosion in NiCr/FLiNaK remains incomplete in two major respects. First, prior atomistic investigations have treated GBs as static, ideal structures, thereby precluding the capture of the structural evolution that the boundary itself undergoes during corrosion, such as local disordering, defect accumulation, and partial dealloying within the GB region \cite{ARKOUB2025113903}. Second, no atomistic mechanism has been established for the dynamic GB migration observed experimentally during molten salt corrosion. Capturing the dynamic GB structural evolution and migration requires an atomistic approach that simultaneously resolves bond-breaking chemistry and accesses the spatial and temporal scales of microstructural rearrangement. Reactive molecular dynamics (RMD) with the ReaxFF formalism \cite{Reaxff1, Reaxff2}, in which dynamically updated bond orders permit bond formation and breaking during the simulation, serves as a feasible method \cite{arkoub2024,arkoub2025}. The approach has been used to investigate a range of metal--environment corrosion processes atomistically, including iron--surface dissolution in chloride electrolytes \cite{shen}, supercritical water \cite{huang}, and aqueous and acidic solutions \cite{farzi}; atmospheric oxidation of iron surfaces \cite{ram} and grain--boundary--mediated mass transport and corrosion in iron \cite{kuan,zigen}. Of particular relevance to the present work, a ReaxFF FLiNaK--NiCr force field has been developed and validated against density-functional reference data for the fluoride--metal corrosion in this system \cite{arkoub2024}.

To fill in the gaps, in this work we apply RMD with the ReaxFF FLiNaK--NiCr force field of Arkoub et al.~\cite{arkoub2024} to compare the corrosion response of four symmetric tilt GBs ($\Sigma3(111)$, $\Sigma11(113)$, $\Sigma5(012)$, and $\Sigma5(013)$), spanning a range of GB energies and surface orientations representative of low--energy boundaries observed in FCC Ni \cite{sangid, chenjie}. Bulk structures with (110) and (100) surfaces serve as references to isolate the surface-orientation contribution from the GB contribution. By tracking metal dissolution, fluorine localization, atomic mobility, and GB migration over a common simulation timescale, we (i) identify a two--regime corrosion response in which surface orientation governs the initial stage and GB character governs the longer--time spatial localization, (ii) establish a corrosion--driven GB migration mechanism, and (iii) identify the coherent $\Sigma3(111)$ twin as a candidate for GB engineering of corrosion-resistant Ni--Cr alloys in molten fluoride environments.

\section{Method}

\subsection{Simulation setup}

\begin{figure}[!ht]
	\centering
	\includegraphics[width=1.0\textwidth]{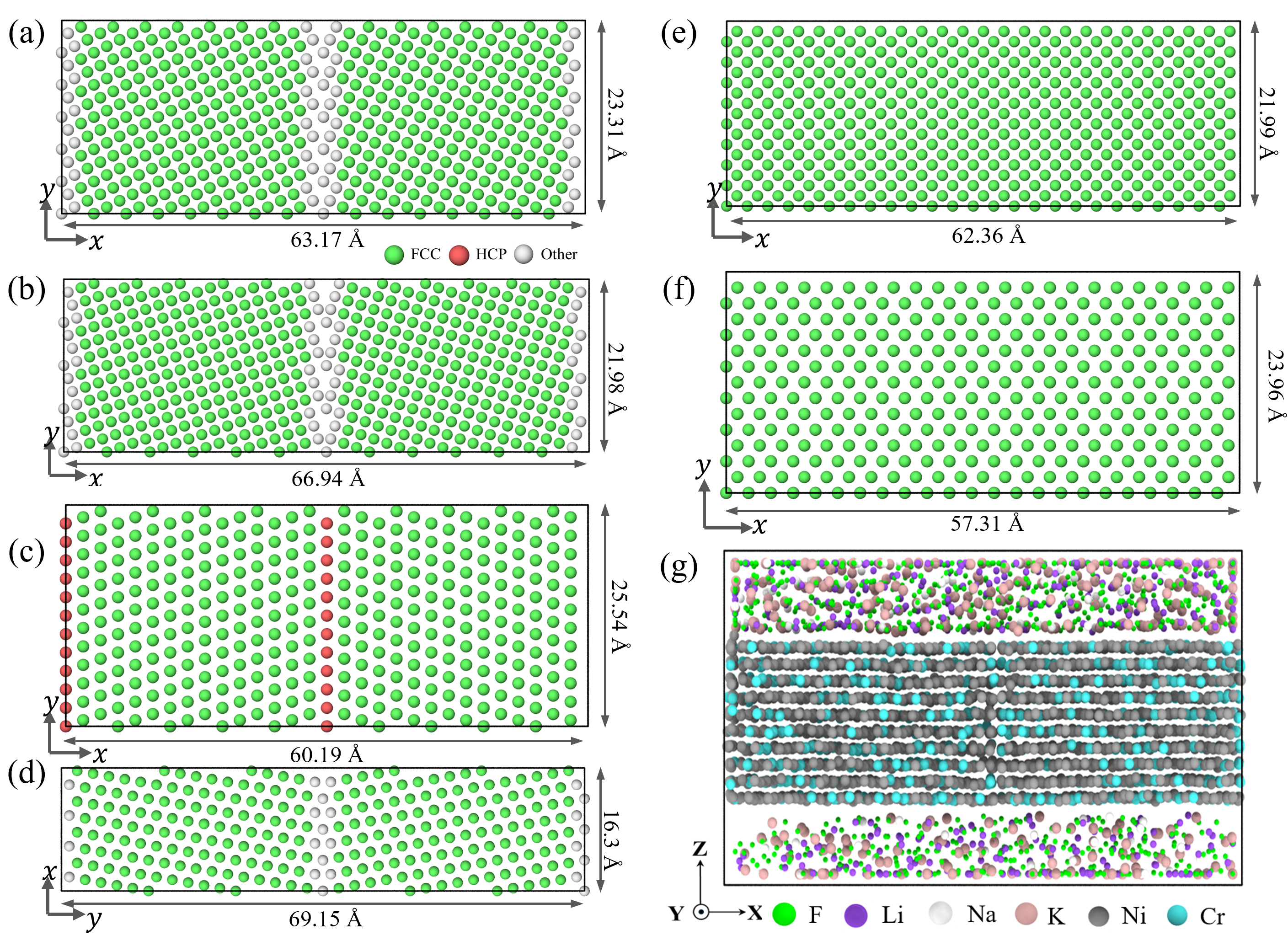}
	\caption{Top view of the structures (a) $\Sigma5(012)$, (b) $\Sigma5(013)$, (c) $\Sigma3(111)$, (d) $\Sigma11(113)$, (e) Bulk(110), and (f) Bulk(100). (g) Representative salt-alloy system.}
	\label{fig:Figure 1}
\end{figure}

RMD simulations were performed using the Amsterdam Modeling Suite (AMS) implementation of ReaxFF \cite{AMS, Reaxff1}, with the FLiNaK--NiCr force field developed and validated by Arkoub et al.~\cite{arkoub2024}. Four bicrystalline Ni$_{75}$Cr$_{25}$ alloy slabs containing symmetric tilt GBs ($\Sigma3(111)$ and $\Sigma11(113)$ with (110) surface orientation, and $\Sigma5(012)$ and $\Sigma5(013)$ with (100) surface orientation) and two reference slabs, Bulk(110) and Bulk(100), were constructed (Figure \ref{fig:Figure 1}). The crystallographic characteristics of the GB models used in this study are summarized in Table S1 of the Supplementary Information (SI). Cr atoms were distributed randomly within each atomic plane ($x-y$ plane) at a fixed concentration of 25\%, yielding a random Ni$_{75}$Cr$_{25}$ solid solution.

Each alloy slab was first equilibrated under the isothermal-isobaric ensemble (NPT) for 50~ps at 800$^{\circ}$C and 1~atm using the Berendsen thermostat and barostat \cite{berendsen}. Molten FLiNaK (LiF--NaF--KF, $46.5{:}11.5{:}42$~mol\%, eutectic composition), which has been commonly studied \cite{frandsen}, was then placed on top and bottom sides of the relaxed slab at a target density of 1.8~g\,cm$^{-3}$ \cite{williams}, representative of FLiNaK at 800$^{\circ}$C. The combined alloy--salt system was subsequently run under canonical ensemble (NVT) at 800$^{\circ}$C for 500~ps with a 0.25~fs timestep and periodic boundary conditions applied in all three directions. A representative alloy--salt cell is shown in Figure~\ref{fig:Figure 1}(g). The structural characteristics and atom counts are summarized in Table~\ref{tab: Table 1}.

\begin{table}[!ht]
\centering
\begin{tabular*}{1.0\textwidth}{@{\extracolsep{\fill}}lllll}
\toprule
Structure &
\begin{tabular}[c]{@{}l@{}}Relaxed Cell Dimensions\\ (with salt) (\AA)\end{tabular} &
\begin{tabular}[c]{@{}l@{}}Salt Molecules\\ (LiF, KF, NaF)\end{tabular} &
\begin{tabular}[c]{@{}l@{}}Metal Atoms\\ (Ni, Cr)\end{tabular} &
\begin{tabular}[c]{@{}l@{}}Total\\ Atoms\end{tabular} \\
\midrule
$\Sigma3(111)[110]$  & 61.89 $\times$ 26.23 $\times$ 40.38 & 400, 361, 99 & 2167, 713 & 4600 \\
$\Sigma5(012)[100]$  & 64.46 $\times$ 24.01 $\times$ 35.64 & 337, 304, 83 & 1809, 591 & 3848 \\
$\Sigma5(013)[100]$  & 68.60 $\times$ 22.53 $\times$ 35.74 & 337, 304, 83 & 1807, 593 & 3848 \\
$\Sigma11(113)[110]$ & 16.70 $\times$ 71.54 $\times$ 40.34 & 294, 266, 73 & 1588, 524 & 3378 \\
Bulk(110)            & 58.03 $\times$ 25.01 $\times$ 39.14 & 328, 296, 81 & 1936, 640 & 3986 \\
Bulk(100)            & 63.82 $\times$ 22.97 $\times$ 36.42 & 324, 292, 80 & 1691, 559 & 3642 \\
\bottomrule
\end{tabular*}
\caption{Simulation cell specifications for Ni--Cr alloy grain-boundary and bulk systems.}
\label{tab: Table 1}
\end{table}

\subsection{Trajectory analysis}

Production trajectories were analyzed using the OVITO package \cite{Ovito}. For each system, 8 independent simulations were performed with different random initial velocity seeds for ensemble averaging. Metal--fluorine and metal--metal bonds were identified at each configuration snapshot, with salt cations (Li, Na, K) excluded. Following prior ReaxFF studies of NiCr corrosion in molten FLiNaK \cite{arkoub2024, arkoub2025}, the metal--fluorine bond cutoff was defined as 
$r_{\mathrm{cut}}^{\mathrm{F-M}} = 0.8\,(r_{\mathrm{vdW}}^{\mathrm{F}} + r_{\mathrm{vdW}}^{\mathrm{M}})$, a Ni or Cr atom was classified as dissolved when its metal-neighbor count within the first--nearest--neighbor distance dropped below three, and a fluorine atom was classified as adsorbed when bonded to two or more non-dissolved metal atoms. These thresholds are selected to be robust against thermal fluctuations and transient metal--salt interactions at 800$^{\circ}$C.

Spatial profiles of dissolved atoms and adsorbed F were constructed by one--dimensional binning along the direction normal to the GB plane. Local F coverage was calculated using a bin width of 0.5~\AA, whereas local metal dissolution profiles employed a bin width of 0.8~\AA, to provide sufficient spatial resolution while reducing statistical noise. In both cases, per--bin quantities were time--averaged over the last 400~ps of the production trajectory. The mean--squared displacement (MSD) of atoms initially in the top alloy layer was computed over the production run; atoms meeting the dissolution criterion were removed from the computation so that the MSD reflects the mobility of atoms remaining bound to the alloy. The corresponding self--diffusion coefficients were then determined from the linear regime of the MSD under the assumption of two--dimensional diffusion. GB positions were tracked based on common neighbor analysis: non--FCC atoms identify the disordered $\Sigma5$ and $\Sigma11$ boundaries, while HCP--classified atoms capture the coherent $\Sigma3(111)$ twin. Atoms within the top and bottom 5\% of the metal cell along $Z$ were excluded to avoid uncertainty from the corroded surface region. For each of the two GBs in the system, the GB position was defined as the mean coordinate of the identified GB atoms along the GB plane normal, and migration was quantified as the displacement of this mean position from its initial value. Additional methodological details are provided in the SI (Equations S1--S7).

\section{Results and Discussion}

\subsection{Preferential Cr dissolution and orientation-dominated initial corrosion}
\begin{figure}[!ht]
	\centering
	\includegraphics[width=1.0\textwidth]{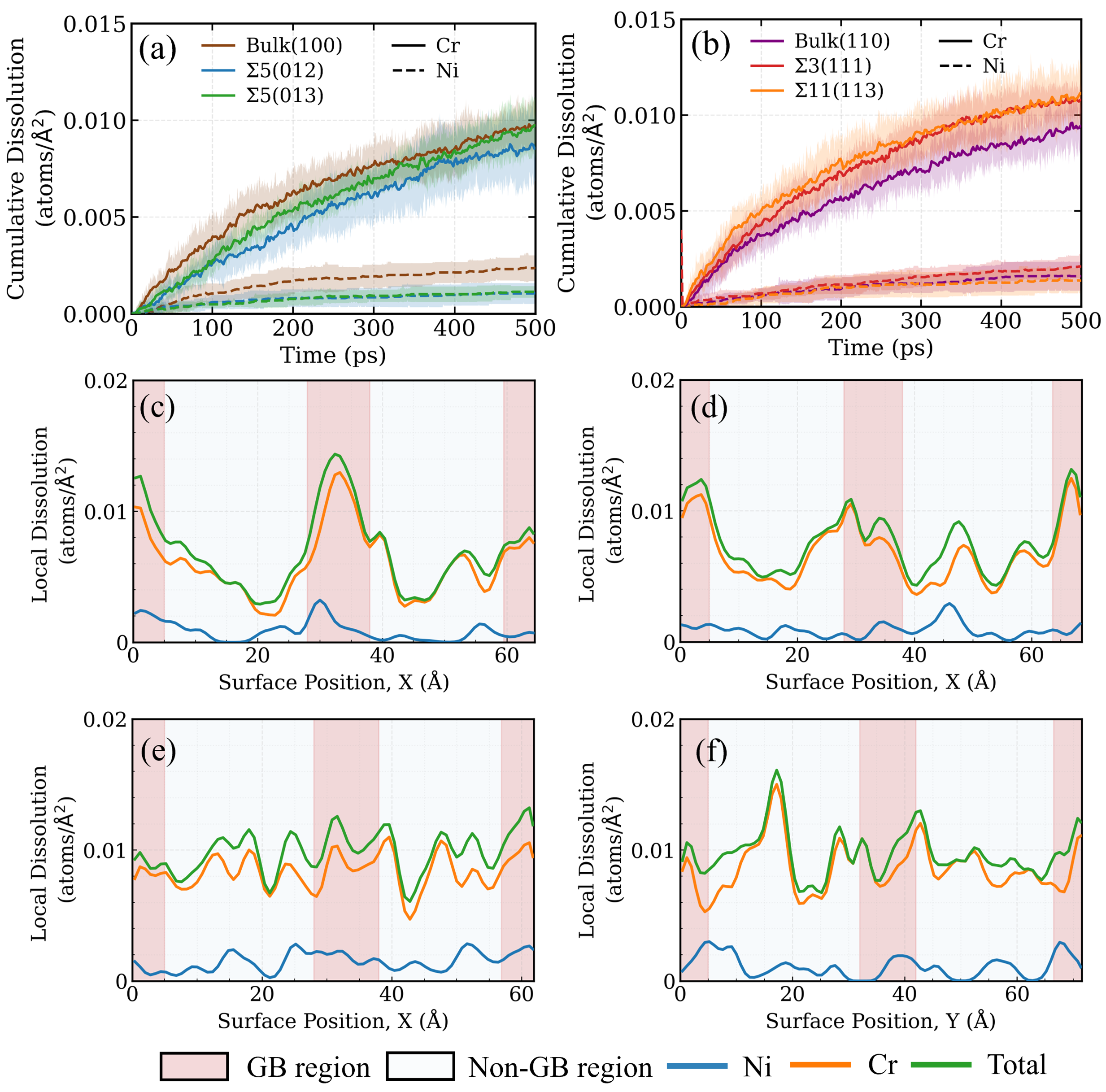}
	\caption{Dissolution of Ni and Cr atoms per surface area for (a) (100) surfaces and (b) (110) surfaces versus time. Shaded regions indicate the standard deviation. Local distribution of dissolved Ni and Cr atoms along the direction normal to the GB plane for (c) $\Sigma5(012)$, (d) $\Sigma5(013)$, (e) $\Sigma3(111)$, and (f) $\Sigma11(113)$. The $\Sigma5$ boundaries exhibit GB--centered dissolution peaks, whereas $\Sigma3(111)$ shows a more uniform dissolution profile, highlighting the role of GB character in controlling corrosion localization.}
	\label{fig:Figure 2}
\end{figure}

Figure~\ref{fig:Figure 2}(a) and (b) compare the dissolution of Ni and Cr from Ni--Cr alloys into molten FLiNaK for the GB--containing structures and the corresponding bulk surfaces. For all structures, Cr dissolution is substantially higher than Ni dissolution, indicating preferential Cr removal during molten salt corrosion. This behavior is consistent with the well-established dealloying tendency of Ni--Cr alloys in molten fluoride salts, where Cr is selectively destabilized at the alloy--salt interface through strong interactions with fluorine-containing species. Previous experimental studies have reported Cr depletion from Ni--Cr alloys exposed to molten fluoride salts, while atomistic simulations have shown that Cr--F bond formation facilitates Cr removal in salt environments~\cite{chan2024, Olson2009MaterialsCI, ARKOUB2025113903, osti_1871749}. As illustrated by the atomic density profiles at the alloy--salt interface (Figure S2 in SI), Cr depletion near the interface is accompanied by the formation of a Ni-rich surface layer, which is consistent with previous simulation and experimental studies \cite{mills2025uncovering,arkoub2024,arkoub2025}.

The comparison also shows that the crystallographic orientation of the exposed alloy surface primarily governs the early dissolution response. In particular, the structures with (110)--type exposed surfaces, including $\Sigma3(111)$, $\Sigma11(113)$, and Bulk(110), exhibit slightly greater Cr dissolution than the (100)--type systems, including $\Sigma5(012)$, $\Sigma5(013)$, and Bulk(100). Although $\Sigma3(111)$ is a coherent twin boundary that is generally considered corrosion resistant in different systems \cite{wang2013structure, Ebrahimi, chenjie}, its total dissolution remains comparatively high in the present study because the alloy surface exposed to the salt has a (110)-type orientation. This trend is consistent with prior ReaxFF simulations of Ni--Cr bulk surfaces in molten FLiNaK, which showed that the (110) surface is more susceptible to corrosion than the (100) surface~\cite{arkoub2025}. The higher reactivity of the (110) surface can be attributed to its more open surface structure and higher density of under-coordinated surface atoms, which promote stronger metal--fluorine interactions and facilitate the detachment of Cr--containing species into the salt. This interpretation is further supported by the atomic density profiles across the alloy--salt interface (Figure S2 in SI). The distance between the first F and Ni density peaks provides a measure of the interfacial separation between molten FLiNaK and the alloy surface. The extracted F--Ni peak separations are approximately 1.6~\AA{} for $\Sigma5(012)$, 1.8~\AA{} for $\Sigma5(013)$, and 1.4~\AA{} for both $\Sigma3(111)$ and $\Sigma11(113)$. The shorter F--Ni separation in the (110)--oriented systems indicates closer fluorine ions to the alloy surface and stronger metal--fluorine interfacial interactions, consistent with their higher initial dissolution tendency.  

This orientation dependence is important for interpreting the GB--containing systems. The overall dissolution curves for GB--containing and bulk systems with the same surface orientation remain relatively close. However, extending the simulations to 1 ns (Figure S1 in SI) reveals a gradual divergence, with both $\Sigma5(012)$ and $\Sigma5(013)$ exhibiting higher Cr dissolution than Bulk(100) after approximately 700 ps. This delayed crossover suggests that the influence of GB character on the overall corrosion rate becomes more pronounced at longer times as GB-assisted corrosion develops. This indicates that the influence of GB character on corrosion rate is not fully captured by total dissolution alone over this timescale. Instead, as discussed below, the more distinctive role of GB character emerges from the spatial distribution of dissolution, corrosion localization, and GB migration. 

\subsection{Effect of GB character on corrosion localization}
The spatial dissolution profiles in Figure~\ref{fig:Figure 2}(c--f) reveal that GB character strongly affects the spatial distribution of dissolution, even when differences in the overall dissolution rate are relatively modest. For $\Sigma5(012)$ and $\Sigma5(013)$, dissolved Ni and Cr atoms are concentrated near the GB plane, with pronounced peaks at the boundary locations and much lower dissolution from the inter--boundary grain--surface regions. This redistribution is analogous to nano--galvanic coupling, with the higher--energy GB acting as a preferential anodic site relative to the ordered surrounding lattice. In contrast, $\Sigma3(111)$ exhibits a nearly flat dissolution profile with no apparent peaks at the boundary, showing that the coherent twin does not function as a preferential corrosion pathway. $\Sigma11(113)$ exhibits mild spatial variations in dissolution but lacks the pronounced GB--centered localization observed for the $\Sigma5$ boundaries, indicating a weaker influence of the GB on corrosion localization.

\begin{figure}[!ht]
	\centering
	\includegraphics[width=0.8\textwidth]{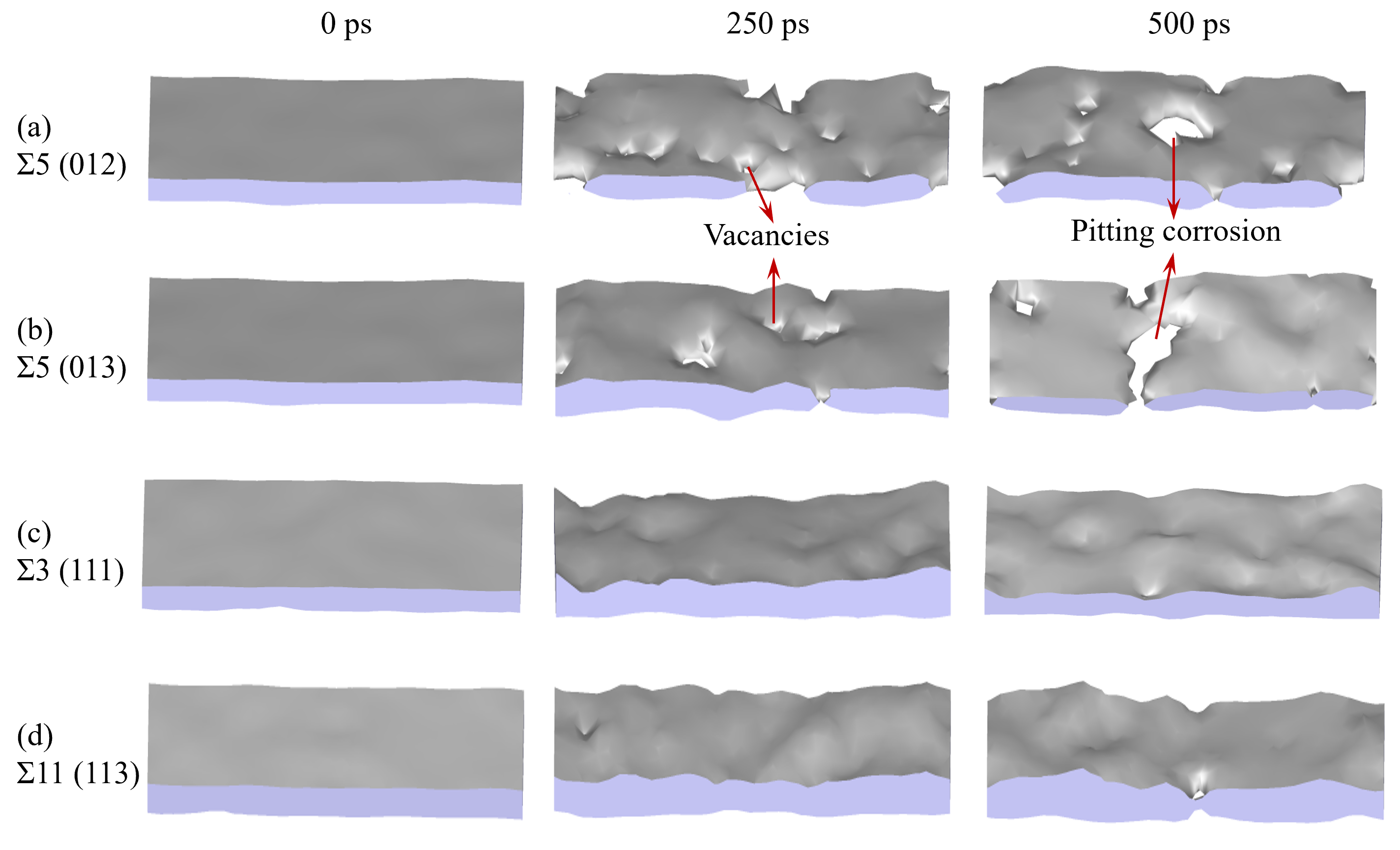}
	\caption{Snapshots of the surface mesh of NiCr slabs with GBs captured at 0 ps, 250 ps, and 500 ps. The gray regions depict the surface mesh, while the purple cross--sections indicate the sides.}
	\label{fig:Figure 3}
\end{figure}

The surface morphology evolution in Figure~\ref{fig:Figure 3} shows that the spatial redistribution of dissolution has direct structural consequences. Initially, all GB--containing slabs exhibit relatively smooth surfaces. During corrosion, the $\Sigma5(012)$ and $\Sigma5(013)$ systems develop vacancies and vacancy clusters near the GB plane by 250~ps,  which evolve into pronounced pit-like recession by 500~ps. Thus, local Cr depletion leaves a vacancy--rich region near the boundary, which promotes further atomic rearrangement and subsequent damage at the boundary. By contrast, $\Sigma3(111)$ and $\Sigma11(113)$ do not show GB--mediated localization, and corrosion produces surface roughening rather than GB--centered pitting. 

\begin{figure}[!ht]
	\centering
	\includegraphics[width=0.65\textwidth]{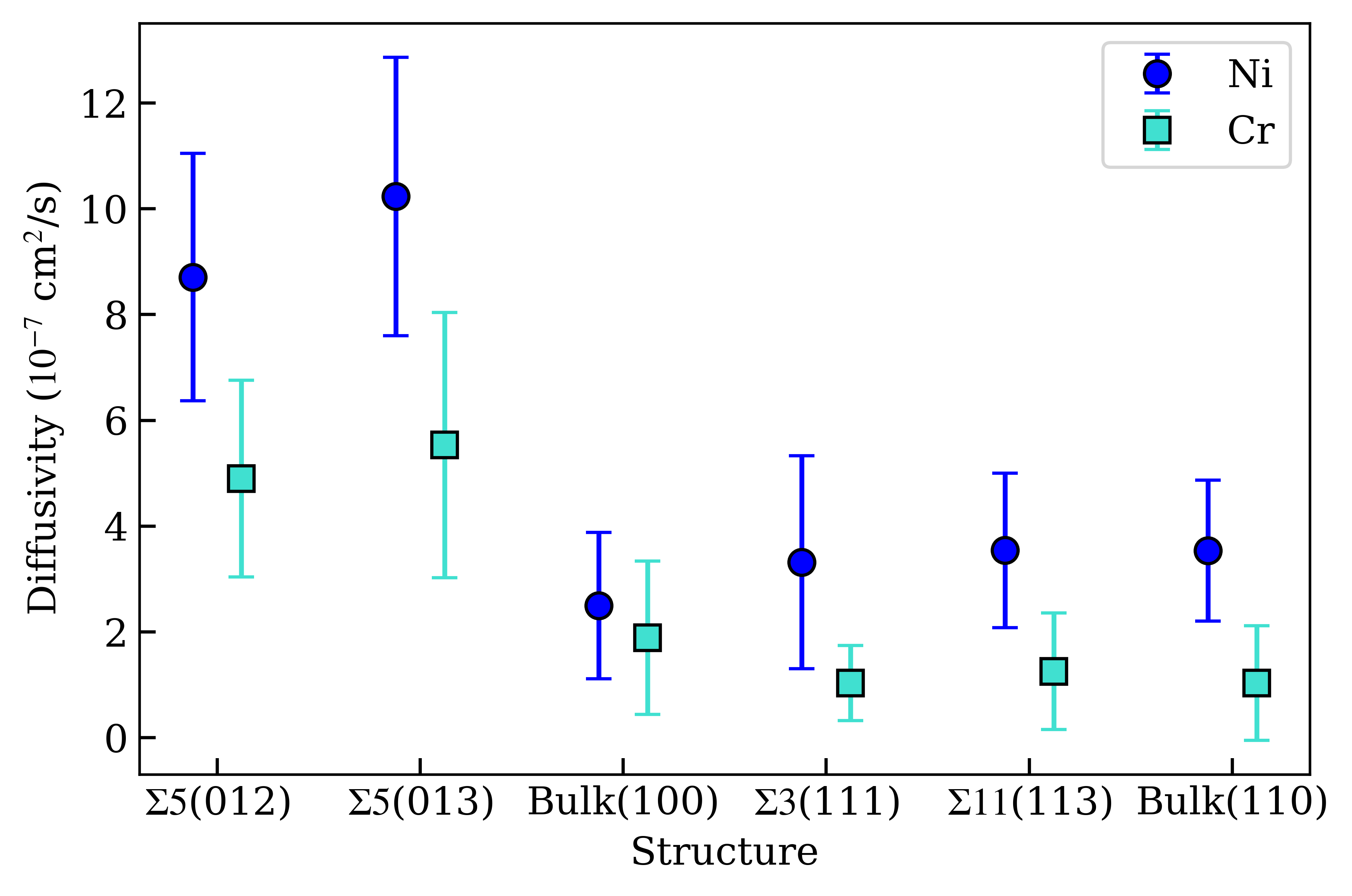}
	\caption{Two--dimensional diffusion coefficients of Ni and Cr atoms in the top layer for all structures at 800$^\circ$C. Error bars indicate standard deviation.}
	\label{fig:Figure 4}
\end{figure}

The morphology evolution can be further understood from the effective surface-layer diffusivities extracted from the MSD slopes (Figure S4 in SI). The MSD analysis tracks atoms initially located in the first alloy layer exposed to molten FLiNaK; therefore, the extracted diffusivities represent an average surface-layer response that includes both GB and inter-boundary regions (Table S2 in SI). From Figure \ref{fig:Figure 4}, the $\Sigma5(012)$ and $\Sigma5(013)$ systems exhibit the highest diffusivity for both Ni and Cr, consistent with their stronger surface recession and pit--like morphology. Ni diffuses faster than Cr in all structures, indicating that the residual Ni-rich surface formed after preferential Cr dissolution remains highly mobile. This result connects to the broader dealloying mechanism. Selective Cr dissolution creates vacancies and local free volume, while fast Ni surface diffusion allows the residual surface to relax, roughen, and coarsen. In classical dealloying, a porous morphology develops through the competition between the selective dissolution of the reactive element and the surface diffusion of the residual noble component~\cite{erlebacher2001evolution}. Recent studies of Ni--Cr corrosion in molten FLiNaK have similarly shown that Cr depletion, point--defect accumulation, Ni redistribution, and pore/channel evolution are coupled during dealloying corrosion \cite{mills2025uncovering,mills2024elucidating,chan2024}. The high surface--layer diffusivity in the $\Sigma5$ systems combined with the localized Cr removal explains the efficient pit--like degradation, whereas the more uniform corrosion of $\Sigma3(111)$ produces less severe localized morphology.

\begin{figure}[!ht]
	\centering
	\includegraphics[width=0.95\textwidth]{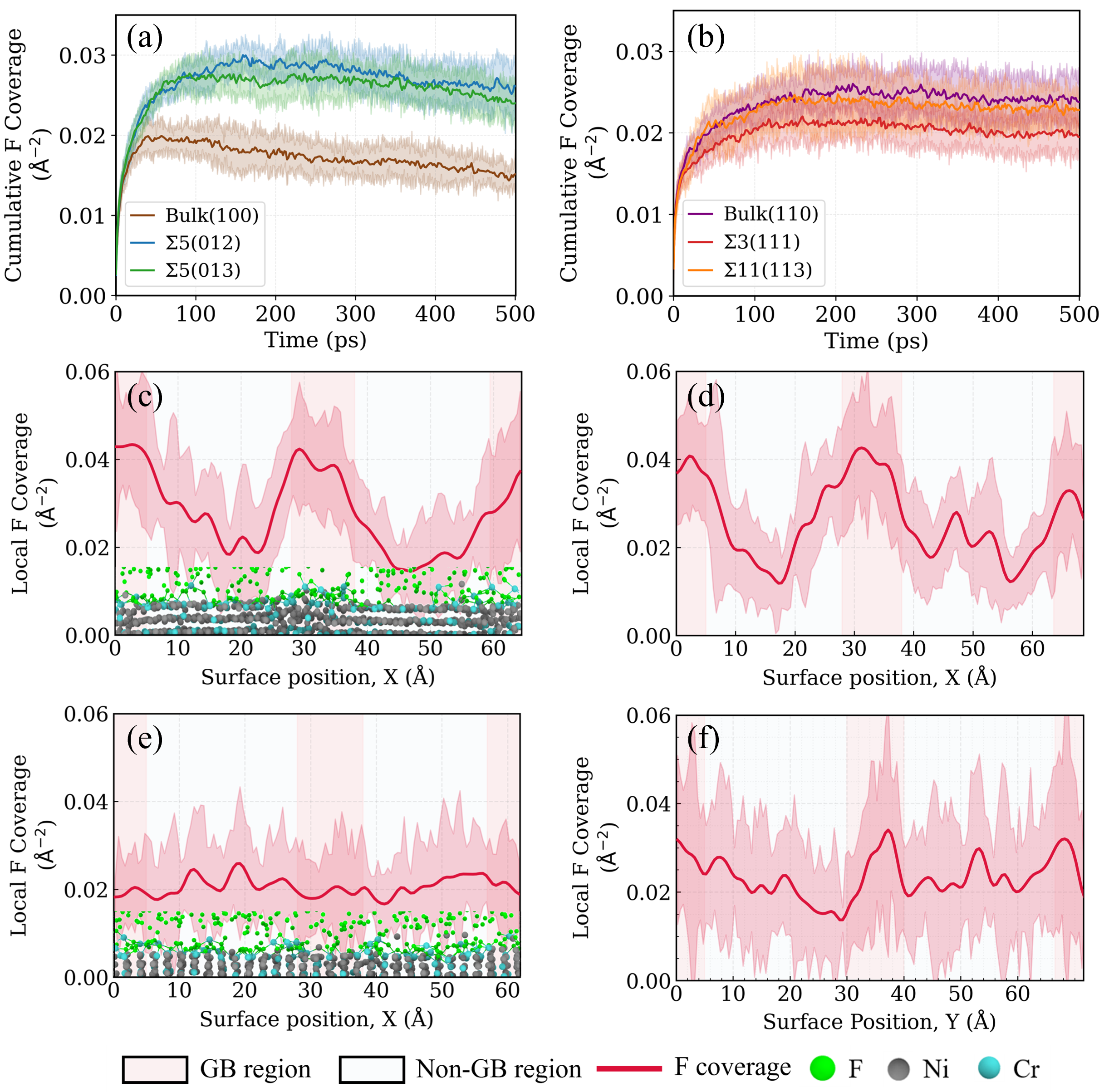}
	\caption{Fluorine coverage per surface area (\text{\AA}$^{-2}$) vs. time (ps) for structures with (a) (100) surface and (b) (110) surface. Local Fluorine coverage per area (\text{\AA}$^{-2}$) along the direction normal to the GB plane for (c) $\Sigma5(012)$, (d) $\Sigma5(013)$, (e) $\Sigma3(111)$, and (f) $\Sigma11(113)$. Shaded bands indicate the standard deviation. The $\Sigma5$ boundaries exhibit GB-centered fluorine coverage peaks, whereas $\Sigma3(111)$ shows a more uniform fluorine coverage. Panels (c) and (e) additionally show atomic configurations for the $\Sigma5(012)$ and $\Sigma3(111)$ systems, selected as representative examples of the (100)-- and (110)--oriented structures, respectively. These snapshots illustrate the contrasting spatial distributions of interfacial F (green), Ni (gray), and Cr (cyan) atoms.}
	\label{fig:Figure 5}
\end{figure}

Fluorine coverage provides a direct probe of the local salt--alloy interaction that precedes Cr dissolution. Because the calculated coverage depends on the cutoff used to identify interfacial F atoms and F--metal bonding, the more reliable information comes from trends within the same orientation. Figure~\ref{fig:Figure 5}(a) and (b) compare the F coverage for the (100)--oriented systems and the (110)--oriented systems, respectively. For all systems, F coverage rises rapidly during the first 50--100~ps and then approaches a quasi-steady level. This rapid saturation indicates that fluorine adsorption occurs early in the corrosion process, before the slower development of pit--like morphology. For the (100)--oriented systems, the $\Sigma5$ structures show higher F coverage than Bulk(100), indicating that the $\Sigma5$ GBs introduce additional adsorption sites beyond those available on the crystalline surface. For the (110)--oriented systems, $\Sigma3(111)$, $\Sigma11(113)$, and Bulk(110) show broadly similar average coverage, indicating that these boundaries add little additional adsorption capacity to an already reactive (110)--type surface. 

The spatial coverage profiles in Figure~\ref{fig:Figure 5}(c--f) identify where the fluorine-driven corrosion chemistry is concentrated. For $\Sigma5(012)$ and $\Sigma5(013)$, F accumulates near the GB plane and decreases toward the inter-boundary grain-surface regions. These F--enrichment peaks coincide with the dissolved-metal peaks in Figure~\ref{fig:Figure 2}(c-f), indicating that local F adsorption is closely coupled to local metal removal. In contrast, $\Sigma3(111)$ shows a nearly uniform F distribution, consistent with its flat dissolution profile and the absence of GB-centered pitting. $\Sigma11(113)$ again exhibits an intermediate response. Representative Cr--F bonding snapshots are shown in Figure S3 in SI. In $\Sigma5(012)$, Cr--F bonds cluster around the GB region, whereas in $\Sigma3(111)$ they are distributed across the exposed surface. Prior DFT and reactive MD studies have shown that F adsorption and Cr--F bond formation promote Cr destabilization and removal from Ni--Cr alloys in fluoride environments~\cite{ARKOUB2025113903,arkoub2024, Yin}. With the GB considered, the $\Sigma5$ boundary concentrates F adsorption and Cr--F bonding into a narrow GB-centered region, while the coherent $\Sigma3(111)$ boundary supports a more spatially distributed reaction across the surface.

\subsection{Corrosion-assisted grain-boundary migration}

\begin{figure}[!ht]
	\centering
	\includegraphics[width=1.0\textwidth]{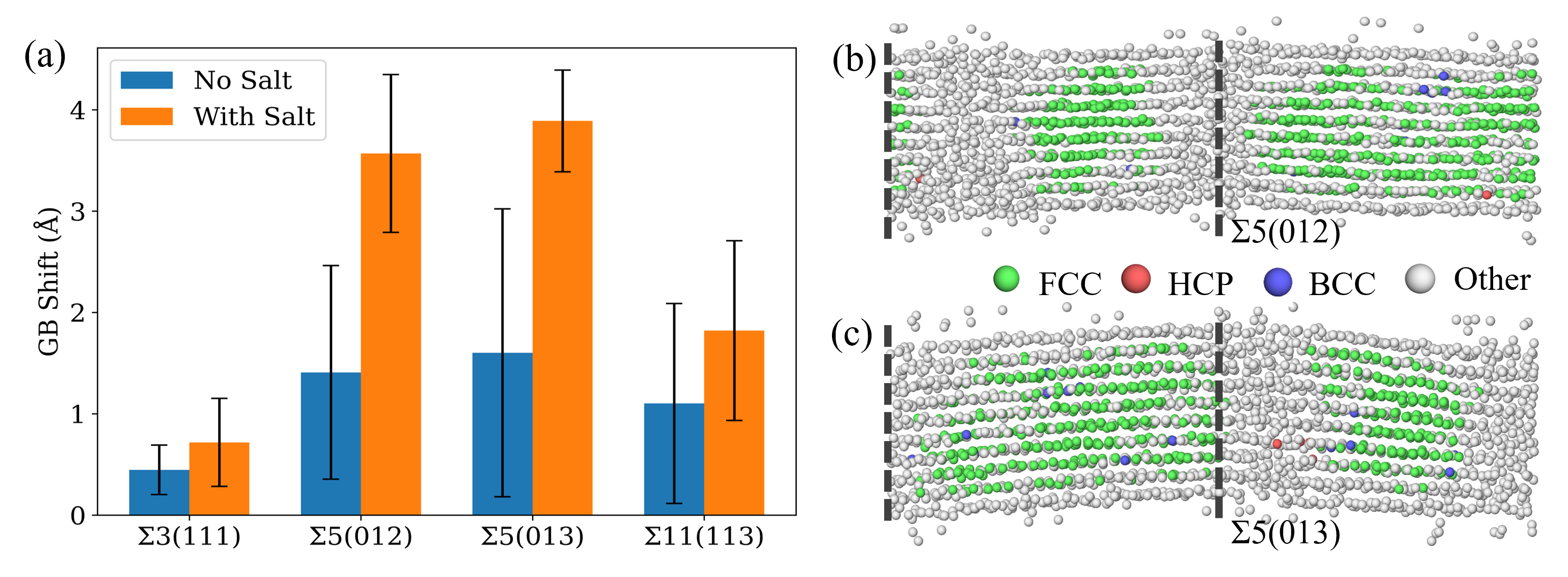}
	\caption{(a) GB shift for the four GB structures with and without molten salt exposure after 500 ps. (b) and (c) Representative snapshots of $\Sigma5(012)$ and $\Sigma5(013)$, respectively, showing the final GB positions after 500 ps. Dashed lines indicate the initial GB positions.}
	\label{fig:Figure 6}
\end{figure}

Corrosion of the GB systems leads to apparent GB migration. Figure~\ref{fig:Figure 6}(a) compares GB displacement with and without molten FLiNaK for all four boundary types. Without salt, GB shifts are modest, with average values ranging from 0.45~\AA{} to 1.60~\AA{}, which stem from thermally activated boundary relaxation and migration. By comparison, salt exposure substantially increases migration, particularly for the high-energy boundaries (e.g., shifts reach $\sim$3.56~\AA{} for $\Sigma5(012)$ and $\sim$3.89~\AA{} for $\Sigma5(013)$) (The data for the GB shift is added in Table S2 of SI). The relative migration trend aligns with Cr dissolution localization, suggesting that the accelerated GB migration is tied to the localized corrosion process, and GB character plays a critical role. Fluorine ions from the salt penetrate the structurally open $\Sigma5$ GB region, as seen from the broad, diffuse F density profile at the $\Sigma5$ interface in contrast to the sharp, surface-confined F peak at $\Sigma3(111)$ (Figure S2 in SI). This penetration is enabled by the excess free volume and structural disorder of high-energy boundaries \cite{sangid, chenjie}. Once localized at the boundary, F forms preferential Cr--F bonds at GB--adjacent sites (Figure S3 in SI), weakening Cr--Ni metallic coordination and further driving selective Cr dissolution \cite{ARKOUB2025113903}. The elevated atomic mobility at $\Sigma5$ boundaries, consistent with the higher MSD reported for GB atoms under molten-salt corrosion conditions \cite{arkoub2026atomistic}, enables rapid redistribution of the resulting vacancies along the GB plane. The accumulated vacancy supersaturation lowers the thermodynamic stability of the near--boundary lattice atoms, driving the boundary toward the undisturbed material. The migrating interface leaves behind a Ni--enriched, Cr--depleted wake that retains the FCC lattice structure (Figures~\ref{fig:Figure 6}(b) and (c)). The Cr--depleted, Ni--enriched composition of this wake zone also accounts for the reduced dissolution observed in the inter-boundary regions of the $\Sigma5$ slabs in Figure \ref{fig:Figure 2}(c) and (d).

This corrosion-enhanced boundary motion is closely related to DIGM. Classical DIGM describes the migration of a GB driven by solute redistribution, where a moving boundary leaves behind a compositionally modified wake as the system reduces chemical or elastic free energy associated with a concentration gradient~\cite{gbshift2,gbshift3,gbshift4}. For example, studies on Ni--based alloys have shown that oxidation couples to Cr depletion, Ni enrichment, and migrated GBs, linking DIGM as a necessary condition for intergranular oxidation and crack-tip degradation~\cite{gbshift,xue2022role,shen2019observation}. In this sense, molten-salt corrosion produces a dealloying-driven form of DIGM, where the salt continuously removes the reactive alloying element, while the GB migrates in response to the evolving local composition and structure. Recent molten salt studies further suggest that DIGM may contribute to the formation of wormhole--like corrosion channels and vacancy supersaturation in Ni--Cr alloys exposed to fluoride salts~\cite{yang2023,teng2025}. The two--dimensional planar GB migration captured here may represent the early-stage precursor to the one--dimensional channel geometry observed experimentally. As the GB migrates toward the intact region and the molten salt acts as a fluorine source and a Cr sink, the DIGM becomes self-sustaining, where intergranular damage could progress into the alloy interior at a rate set by boundary mobility and local GB structure. The atomistic mechanism identified here provides a microscopic explanation for the experimentally observed coupling between preferential Cr transport and DIGM during molten salt corrosion of Ni--Cr alloys \cite{walter2026}.

The occurrence of DIGM is GB character dependent, as also reported in other corrosion systems \cite{nguejio2017diffusion,shen2019observation}. In the current simulations, $\Sigma$5 GBs exhibit much stronger migration, while the coherent $\Sigma3(111)$ boundary suppresses the coupling between fluorine chemistry and GB motion. The coherent twin boundary has near-zero excess volume at the GB plane \cite{wang2013structure}. Preferential F--Cr interaction and Cr dissolution at the GB plane are consequently minimized, which reduces the driving force for GB motion. Hence, the twin boundary remains structurally stable throughout the simulation despite the appreciable surface dissolution on its (110) --faces. This result clarifies an important aspect of GB character dependence in DIGM susceptibility. The geometric accessibility (the capacity of the boundary to allow preferential F--Cr interaction and F penetration) likely controls whether the corrosion--driven DIGM sequence is activated. It is thus believed that activating this degradation pathway requires two simultaneous conditions: sufficient fluorine activity in the salt to drive Cr dissolution, and sufficient structural openness at the boundary to concentrate that dissolution at the GB plane. The latter suggests a GB engineering strategy via increasing the fraction of coherent low-energy boundaries; this would suppress the F$^-$ penetration at the GB that initiates the corrosion-driven DIGM. This broadly aligns with the prior notion of enhancing intergranular corrosion resistance through GB engineering structurally--ordered low $\Sigma$ GBs in Ni--based and other alloy systems \cite{lin1995influence,chen2021study,xia2011appling}. 

\section{Conclusion}

In this work, RMD simulations were used to investigate how GB character influences molten FLiNaK corrosion of Ni--Cr alloys. Four representative GBs, $\Sigma3(111)$, $\Sigma11(113)$, $\Sigma5(012)$, and $\Sigma5(013)$, were compared with corresponding bulk surface models to separate the effects of surface orientation and GB structure. Across all systems, Cr dissolves preferentially over Ni, confirming that corrosion proceeds through selective Cr removal. The overall dissolution response is strongly influenced by surface orientation, with the (110)--oriented systems showing higher Cr dissolution than the (100)--oriented systems. The primary effect of GB character is the spatial localization of corrosion. The high--energy $\Sigma5(012)$ and $\Sigma5(013)$ boundaries with greater excess volume concentrate F adsorption and metal atom dissolution near the GB plane, while the surrounding grain-surface regions show reduced dissolution. This localized attack develops into vacancy accumulation and pit-like surface recession near the GB. In contrast, the coherent $\Sigma3(111)$ boundary shows a more uniform dissolution profile and limited GB--centered damage, even though its exposed (110)--type surface is intrinsically reactive. The localized dealloying at $\Sigma5$ boundaries substantially enhances their migration relative to no-salt controls, while $\Sigma3(111)$ remains comparatively stable. This provides atomistic evidence for a dealloying--driven form of DIGM, in which fluorine--assisted Cr dissolution and atomic diffusion create a Cr--depleted, vacancy--rich region that drives boundary motion away from the corroded zone. These results reveal a coupled degradation pathway in which susceptible GBs concentrate fluorine chemistry, localize Cr removal, and migrate away from the dealloyed zone. Increasing the fraction of coherent special boundaries, particularly $\Sigma3(111)$ twins, may suppress localized dealloying and corrosion-assisted intergranular migration in Ni--Cr alloys under molten fluoride conditions.

\section*{Author Contributions}
Miaomiao Jin conceived and supervised the research. Sadia Khan performed the calculations, analyzed the data, and drafted the manuscript. All authors interpreted the results, revised, and approved the manuscript.

\section*{Declaration of competing interest}
The authors declare no competing financial or non-financial interests.

\section*{Data availability}
The data that support the findings of this study are available from the corresponding author upon reasonable request.

\section*{Acknowledgments}
This work was supported by the National Science Foundation (NSF) under CAREER Award No. 2340019. Any opinions, findings, conclusions, or recommendations expressed in this material are those of the authors and do not necessarily reflect the views of NSF.

\bibliographystyle{elsarticle-num}
\bibliography{references}

\end{document}